**Electronic structure and energy level schemes of $RE^{3+}$:LaSi$_3$N$_5$ and $RE^{2+}$:LaSi$_3$N$_{5-x}$O$_x$ phosphors (RE= Ce, Pr, Nd, Pm, Sm, Eu) from first principles**

Ismail A. M. Ibrahim[a], Zoltán Lenčéš[a], Pavol Šajgalík[a], Lubomir Benco[a,b,*], Martijn Marsman[b]

[a]*Institute of Inorganic Chemistry, Slovak Academy of Sciences, Bratislava, Slovakia*
[b]*Faculty of Physics and Center for Computational Materials Science, University of Vienna, Vienna, Austria*
[*]corresponding author

**Abstract**

First principles calculations of rare earth (RE)-doped $LaSi_3N_5$ host lattice are performed to obtain the electronic structure, the band gap (BG), and the character of electronic transitions. Doping with both trivalent and bivalent RE cations is inspected. RE 4f states form two bands of occupied and unoccupied states separated by ~5 eV. In $RE^{3+}$-doped compounds 4f states are shifted by ~6 eV to more negative energies compared with $RE^{2+}$-compounds. This stabilization causes that $RE^{3+}$ 4f bands are in a different position relative to the valence band and the conduction band than $RE^{2+}$ 4f bands and therefore different electronic transitions apply. BG of $RE^{3+}$-compounds decreases from ~4.6 eV (Ce) to ~0.5 eV (Eu). Except for $Ce^{3+}$, exhibiting the 4f→5d transition, other $RE^{3+}$-compounds show the charge transfer of the p → 4f character. BG of $RE^{2+}$-compounds increases from ~0.80 eV (Ce, Pr) to ~0.95 eV (Nd, Pm), ~1.43 eV (Sm), and ~3.28 eV (Eu) and the electronic transition is of the 4f→5d character. The energy level scheme constructed from ab initio calculated electronic structures agrees well with the experimental energy level diagram. The agreement demonstrates the reliability of the hybrid functional HSE06 to describe correctly bands of nonbonding RE 4f electrons.

**Introduction**

The past half century has witnessed the discovery of many new luminescent materials and numerous advances in the understanding of basic physical processes governing the operation of inorganic phosphors [1]. Most of luminescent materials are oxides, nitrides, sulfides, oxynitrides and oxysulfides doped with rare-earth (RE) cations. It is well known that the spectroscopic properties of RE cations can be divided to two categories [2]. The first category is the band emission from the 5d→4f transition (e.g. $Eu^{2+}$, $Ce^{3+}$). Because the empty 5d states are not shielded from the crystal field, these transitions can strongly depend on the host lattice [3]. The second category is the charge transfer band emission (e.g. $Pr^{3+}$, $Sm^{3+}$, $Eu^{3+}$), which can be ascribed to the transfer of an electron from 4f states of the RE cation to atoms of the framework. Because 4f electrons are shielded by 5s and 5p electrons, only a weak influence from the host lattice is introduced on the charge transfer transitions [3]. The luminescence properties of all $RE^{3+}$ and $RE^{2+}$ cations in a certain host can be related to each other according to the energy level diagram originally developed by Dorenbos [4-8]. This diagram, showing the energy of the 4f and 5d states of the $RE^{3+}$ and the $RE^{2+}$ cation with respect to the valence (VB) and conduction bands (CB), uses luminescence and optical spectroscopy data. The data are compiled for more than 300 different compounds and the scheme allows the prediction of the 4f and 5d energies of the $RE^{3+}$ and $RE^{2+}$ cations.

Solids doped with RE cations have been subject of many ab initio studies. Most of them either use the cluster model or band-structure approaches based on density functional theory (DFT). The deficiency of the cluster model is that it cannot give the position of the VB and the CB [9]. A number of trivalent lanthanides using DFT-based band-structure calculations were studied by Schmidt et al. [10, 11]. It was observed that within the local density approximation (LDA) and generalized density approximation (GGA) to DFT the self-interaction error associated with the localized nature of 4f electrons prohibited the calculation of the correct position of the 4f band. The shortcomings of standard DFT may be overcome by different approaches. Using the correction of the self-interaction of f electrons Petit et al. [12] found electronic configurations of the RE cation in $REO_2$ and $RE_2O_3$ oxides (RE = Ce, Pr, Nd, Pm, Sm, Eu, Gd, Tb, Dy, Ho) in good agreement with experiments. Another way to improve the standard DFT is to modify the intra-atomic Coulomb interaction through the LDA+U approach [13-15]. It was shown that this approach allowed for a correct treatment of electronic states in $Ce_2O_3$ [16] and in $CeO_2$ [17]. The physical idea behind LDA+U or GGA+U scheme comes from Hubbard Hamiltonian. In the practical implementations, the on-site two-electron integrals are expressed in terms of two parameters. These are the Hubbard parameter $U$, which reflects the strength of the on-site Coulomb interaction, and the parameter $J$, which adjusts the strength of the exchange interaction. In somewhat simplified rotationally invariant method of Dudarev et al. [18] these two parameters are combined into a single parameter $U_{eff} = U - J$. In short, DFT+U correction acts to reduce the one-electron potential locally for the specified orbitals of the respective atoms, therefore reducing the hybridization with orbitals of the host framework. The $U_{eff} = 0$ case represents the DFT limit. Canning et al. [19, 20] used DFT+U method in extensive study of luminescence in Ce-doped scintillators. Tuning the gaps of a series of Ce-doped compounds they derived the parameter $E_{eff} = 2.2$-$2.5$ eV and applying to about 100 new materials they developed a list of candidate materials for new $Ce^{3+}$-activated scintillators.

More elaborate ways to calculate improved thermodynamic and electronic properties of molecules and solids is the use of hybrid functionals and the GW method. Becke has shown that the improvement is unlikely unless exact-exchange information is considered [21]. The calculation of the nonlocal exchange to be used in hybrid functionals, however, drastically increases the computational demands especially for systems with metallic character. To circumvent this bottleneck, Scuseria et al. [22, 23] has developed a new hybrid density functional based on the screened Coulomb potential for the exchange interaction, which enables fast and accurate hybrid calculations. Sautet et al. [24] compared the band gaps and dielectric constants of a series of photovoltaic materials calculated by different functionals and reported that hybrid functional HSE06 always obtained the best agreement with experiment. The most suitable approach for studying single-particle excitation energies of extended systems is the Green function method (GW) [25]. Here, the states are considered as quasiparticle excitations to the electronic ground state, as described by appropriate Green functions. The equation of motion of the single-particle Green function yields the quasiparticle electron and hole states that define the band structure of a system. The band gaps by the GW method agree well with experimental results, therefore the GW method is becoming a standard tool for predicting quasiparticle band structures [26].

In our recent papers we have presented the experimental excitation and emission spectra of $LaSi_3N_5$:RE (RE = Eu, Ce, Sm) phosphors [27-29]. In the computational part of these papers we focused on the location of the band of 4f states between the VB and the CB. The location shows the character of electronic transition and determines the band gap. Low concentration of the RE cation in the $LaSi_3N_5$ host lattice (1 – 6%) required the use of a large supercell. Because the size of the supercell prohibited the use of computationally demanding many-body GW method, we used the hybrid functional HSE06 to localize 4f states in Eu-

[27], Ce- [28], and Sm-doped [29] LaSi$_3$N$_5$ host lattice. Our model doping considered both RE$^{3+}$ and RE$^{2+}$ cations. It was observed that 4f states form two bands of occupied and unoccupied states. All calculated band gaps reasonably compared with experimental data except for first excitation band in Ce-doped phosphor. The excitation energy of the intense band at ~3.5 eV was too small to be compared with the gap of ~4.7 eV calculated for Ce$^{3+}$ cation, and too large compared with ~0.9 eV of the Ce$^{2+}$ cation. Is there some another mechanism responsible for the luminescence in Ce-doped materials? It is generally accepted that in Eu-doped materials responsible for the electronic transition is the Eu$^{2+}$ cation. This cation is more stable than Eu$^{3+}$ because a magic configuration 4f$^7$ is formed. Similar situation is possible with Ce-doped material. Here, the magic configuration 4f$^0$ is possible, if the cation increases its oxidation state to Ce$^{4+}$. A model doping with the Ce$^{4+}$ cation, which was not investigated yet, will be the subject of our future research.

Results on three RE-doped materials (RE = Eu, Ce, Sm) [27-29] show that the position of the 4f band changes with both atomic number and the oxidation state of the RE cation. In this work we compare the electronic structure of the whole series of LaSi$_3$N$_5$:RE phosphors (RE = Ce, Pr, Nd, Pm, Sm, Eu). The aim of the present work is not an extremely accurate calculation of the 4f and 5d level position, but to determine the position of these two bands relative to the VB and CB. Two oxidation states RE$^{3+}$ and RE$^{2+}$ are considered, electronic configuration of which are supposed to play a major role in optically active electron transitions. We therefore compare the electronic configurations and electronic transitions of the RE$^{3+}$ and RE$^{2+}$ cations. From the variation of the band position with increasing atomic number of the RE atom we will construct the energy level scheme and compare the ab initio scheme with the scheme of Dorenbos et al. [8] derived from experimental data. The composition of analyzed samples shows that in RE-doped LaSi$_3$N$_5$ the concentration of O atoms is higher than that corresponding to the lower oxidation state RE$^{2+}$. Such extra oxygen atoms compensate the vacant cationic positions in the host framework. In our previous works on Ce-, Sm-, and Eu-doped phosphors [27-29] we have shown that the formation of vacancies makes only negligible influence of the band gap of the system. In the present work the formation of vacancies is therefore not considered.

**Structure and Computational details**

The structure of the host lattice LaSi$_3$N$_5$ is reconstructed according to the crystallographic data published by Inoue et al. [30] and Hatfield et al. [31]. The nitridosilicate framework consists of SiN$_4$ tetrahedra linked by sharing corners to form rings of tetrahedral units. The lanthanum cation is located in the center of pentagonal space. The lattice vectors $a$ = 7.833 Å, $b$ = 11.236 Å, and $c$ = 4.807 Å form a cell containing 36 atoms (four formula units). The unit cell and the location of the cation in the cavity are displayed in Fig. 1. The distance of the cation to surrounding N atoms varies between ~2.5 Å and ~2.9 Å.

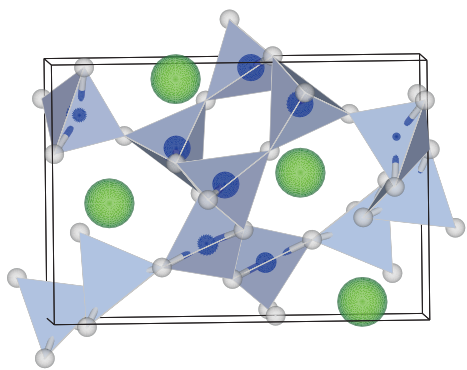

**Fig. 1.** Unit cell of $LaSi_3N_5$. $La^{3+}$ cation is located in the cavity surrounded by $SiN_4$ tetrahedra of the framework (La – large green, Si - blue, N – small gray).

The composition of the RE-doped compound is $La_{1-z}RE_zSi_3N_{5-z}O_{1.5z}$ with $z$ values between 0 and 0.063. To mimic the realistic concentration of RE in $LaSi_3N_5$ a 2×1×2 supercell is formed containing 144 atoms. The $RE^{3+}$-doping is performed via the substitution of one $La^{3+}$ cation with the $RE^{3+}$ cation. In the $RE^{2+}$-doping the $RE^{2+}$ cation replaces one $La^{3+}$ cation and one N atom in the vicinity of the $RE^{2+}$ cation is replaced with the O atom. The oxygen atom delivers to the framework one more electron compared with the N atom. This causes the decrease of the valency of the $RE^{3+}$ cation to $RE^{2+}$. The substitution of one La atom with RE in the supercell results in the composition $La_{0.9375}RE_{0.0625}Si_3N_{4.9375}O_{0.0625}$, very similar to the composition of experimentally prepared material.

The Vienna Ab initio Simulation Package (VASP) [32-34] was used for spin-polarized GGA PW91 [35] calculations. The projector-augmented-wave-function (PAW) approach, developed by Blochl [36] and adapted and implemented in VASP [37] was used for the description of electronic wave functions. An energy cutoff of the plane-wave expansion was 400 eV. The sampling of the Brillouin-zone was restricted to a single point (gamma point). Atomic positions of all atoms were relaxed in the supercell of fixed shape as determined by X-ray structure determination [30, 31]. The atomic relaxation used a stopping criterion of 10-5 eV for the self consistency loop and 10-4 eV for the optimizer. The cell volume was determined using the fit of the energy/volume curve to the Birch-Murnaghan equation of state [38, 39]. The band gap of both $RE^{3+}$ and $RE^{2+}$-doped phosphors was calculated using the

HSE06 functional [22, 23]. For the mixing constant *a* and screening parameter ω the values of 0.25 and 0.207 [40], respectively, were taken.

**Results and discussion**

*4f states in $RE^{3+}$ and $RE^{2+}$*

A relation between the oxidation state of the RE cation and the electronic structure is illustrated on the example of the Sm-doped compound. Figure 2 shows the total and projected DOS of $LaSi_3N_5$ doped with $Sm^{3+}$ and $Sm^{2+}$ in the energy range covering the VB and the CB.

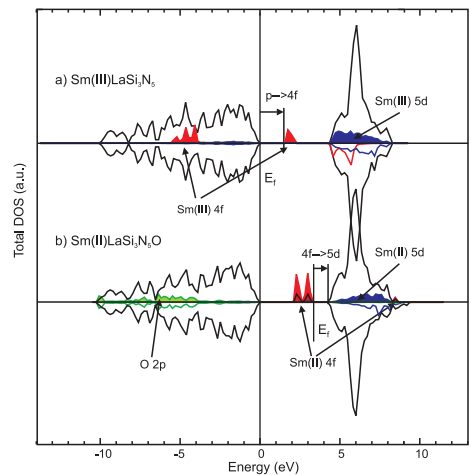

**Fig. 2.** Electronic structure of $LaSi_3N_5$ doped with $Sm^{3+}$ (a) and $Sm^{2+}$ (b) cation. Total DOS is shown in black, Sm 4f character states in red, Sm 5d character in blue, O 2p character states are shown in green.

With the change of the oxidation state of the RE cation no pronounced change of the shape or the position of the VB and the CB relative to each other is observed. The separation of the VB and the CB in both $RE^{3+}$- and $RE^{2+}$ phosphor is ~4.5 eV [29]. The O/N substitution in the $RE^{2+}$ compound introduces an admixture of O 2p states into the VB (Fig. 2b). The O 2p states range mainly between ~–10 eV and ~–4 eV. The location deep in the VB indicates a strong bonding character and means that O atoms in $RE^{2+}$-doped phosphors does not influence optical properties of the substance.

The calculated DOS show that 4f states in both $Sm^{3+}$ and $Sm^{2+}$ remain nonbonding, exhibiting no hybridization with orbitals of atoms of the host framework. The position and shape of the band of 4f states is determined by the crystal field splitting of the $LaSi_3N_5$ host framework. In $Sm^{3+}$ compound spin-up states are collected in two bands (Fig. 2a). More stabilized are the occupied states, which form a triple band located deep in the VB and centered at ~–4.5 eV. This band corresponds to the high-spin configuration $4f^5$. Unoccupied 4f spin-up states, corresponding to two electrons, reside directly in the gap between the VB and the CB at ~2 eV. The splitting of occupied and unoccupied spin-up states is ~6.5 eV. Note that all 4f bands are relatively narrow. This indicates the nonbonding and local character of both occupied and unoccupied 4f states in the $Sm^{3+}$ cation. 4f spin-down states are destabilized compared with the spin-up states and located in the bottom of the CB at ~5 eV. In $Sm^{3+}$-doped $LaSi_3N_5$ the nonbonding 4f electrons are considerably stabilized and localized deep in the VB at ~–4.5 eV. Least stabilized are nonbonding electrons at the Fermi level which are electrons of N 2p lone pairs. The minimum energy electronic transition is therefore valence band charge transfer of the nonbonding electron from N 2p lone pair to the empty band of Sm 4f states localized in the gap between the VB and the CB (Fig. 2a).

In $Sm^{2+}$ compound the spin-up states are collected in two bands, as well (Fig. 2a). Compared with the $Sm^{3+}$ compound (Fig. 2a) both bands are shifted up by ~6 eV. Larger stabilization of 4f states in the $Sm^{3+}$ cation is due to larger Coulomb attraction of electrons in more positively charged cation. This shift in $Sm^{2+}$ doped compound agrees well with the experimental f→d transition energies of divalent lanthanides observed in inorganic compounds [6]. Due to the destabilizing shift the band of occupied 4f states is located directly in the gap. With the decrease of the charge $Sm^{3+} \to Sm^{2+}$ the occupation of the 4f band increases from $4f^5$ to $4f^6$. The band of unoccupied 4f spin-up states moves from ~2 eV in $Sm^{3+}$ to the position at ~8 eV relative to valence band maximum. The band of unoccupied spin-down states moves to more positive energies by ~6 eV, as well. In $Sm^{2+}$ nonboding 4f states are the least stabilized electrons located in the gap below the CB. Because the bottom of the CB is dominated by Sm 5d empty states, the minimum energy transition is intra-atomic 4f → 5d electron transfer. In our recent work we presented the experimental gap of $LaSi_3N_5$:Sm phosphor of 2.12 eV [29] and compared the experimental gap with values calculated for $Sm^{3+}$- and $Sm^{2+}$-compounds. The experimental gap of 2.12 eV compared well with 2.01 eV calculated for $Sm^{3+}$-doped compound and the corresponding transition was valence band charge transfer p→4f.

*Total DOS*

The electronic structure of stoichiometric $LaSi_3N_5$ and a series of $LaSi_3N_5$ host compounds doped with trivalent $RE^{3+}$ cation (RE = Ce, Pr, Nd, Pm, Sm, Eu) is shown in Fig. 3. The spectrum of energy levels of the host $LaSi_3N_5$ compound consists of the *s* band at –34 eV, *sp* band ranging between –14 eV and –20 eV, the valence band between –10 eV and 0 and the conduction band between 4 eV and 7 eV. The shape and the composition of bands, and their relative position are driven by bonding between atoms of the host framework. The structure of LaSi$_3$N$_5$ consists of covalently bound Si-N framework and La$^{3+}$ cations residing in cavities. The ionic character of bonding of the cations causes that bands originating from La atoms are deep lying and narrow. La 5s states form the narrow band at –34 eV and La 5p states contribute to the middle part of the *sp* band at ~–17 eV. For the sake of simplicity, Fig. 3 displays not all details on the composition of bands. A strong covalent bonding between Si and N in the host framework causes that the valence band is as broad as 10 eV. The strongest Si–N bonding produces states at –10 eV. On the contrary, states of nonbonding electrons, residing on N as lone electron pairs, form states at the edge of the VB at 0 eV.

Upon the substitution of the La$^{3+}$ cation with RE$^{3+}$ cation the shape and the location of bands of the host compound remain practically unchanged. Both the 4f and 5d states of the RE dopant are of nonbonding character and form bands in the vicinity of the band gap. In all doped compounds 5d states are empty and spread over the whole width of the CB. The center of mass of the 5d states is located at the bottom of the CB at a rather constant position and invariant with the type of lanthanide ion. This agrees well with the conclusion drawn from the analysis of experimental data [5, 8]. In the energy level scheme constructed from experimental data, however, 5d energy levels of trivalent RE cations are localized slightly below the CB [8]. The location of the 5d states within the CB can be ascribed to a drawback of the computational procedure, which optimizes the position only for occupied states. How is the position of an unoccupied level improved upon use of the hybrid functional, applied in this work, is not investigated yet. First attempts to address this problem at the DFT+U level have done Canning et al. [19] who calculated (Ce$^{3+}$)* excited states produced by the transfer of the electron to the Ce 5d state in a number of inorganic compounds.

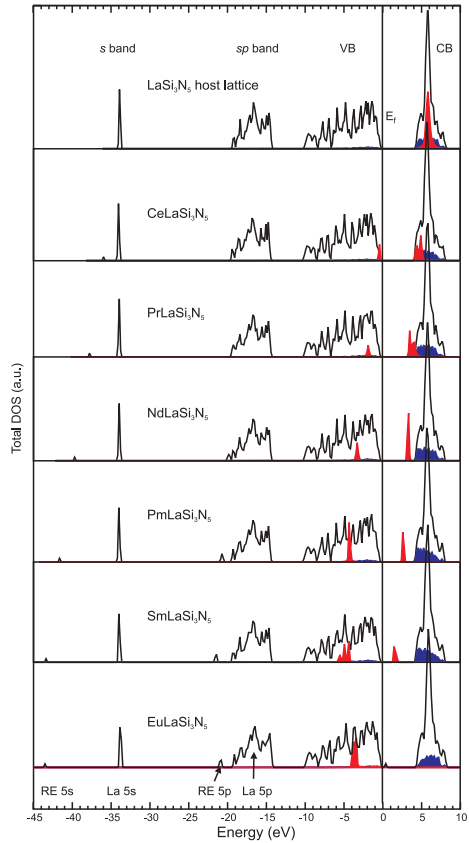

**Fig. 3.** Electronic structure of $LaSi_3N_5$ and $RE^{3+}$-doped $LaSi_3N_5$ (RE = Ce, Pr, Nd, Pm, Sm, Eu). Fermi level is set to 0, total DOS is shown in black, RE 4f character states in red and RE 5d states are shown in blue. RE 4f and 5d states are magnified.

As already shown on the example of the $Sm^{3+}$-doping in Fig. 2, 4f states of the lanthanide atom form two bands of occupied and unoccupied states in all $RE^{3+}$-doped compounds. States of both band are nonbonding, i.e. there is no mixing between 4f orbitals and orbitals of the host lattice. Bands of 4f states therefore remain narrow. The band of occupied states corresponds to the total number of 4f electrons. In the $Ce^{3+}$-compound the electronic configuration is $4f^1$. With increasing atomic number of the RE cation the number of 4f electrons increases to 6 in the $Eu^{3+}$-compound. The separation of the two 4f bands is approximately 5-7 eV for the series $Ce^{3+}$ to $Sm^{3+}$ compounds and decreases to ~4 eV in $Eu^{3+}$ In $Ce^{3+}$ the position of the couple of 4f bands coincides with the top edge of the VB and the bottom edge of the CB (Fig. 3). The $Ce^{3+}$ doped compounds thus can exhibit 4f-5d transition. This is in contrast to other $RE^{3+}$-compounds in which the 4f-5d transition is less probable (cf. below).

With increasing atomic number all energy levels of the RE cation move to more negative values (Fig. 3). The RE 5s band shifts from ~–35 eV ($Ce^{3+}$) to –~44 eV (Eu3+) and RE 5p band from ~–19 eV ($Ce^{3+}$) to –~21 eV ($Eu^{3+}$). The couple of RE 4f bands moves to more negative energies, too. In $Ce^{3+}$ ($4f^1$) the occupied 4f band resides at the Fermi level (Fig.

3). With increasing number of 4f electrons the band moves continuously to –4.5 eV in $Sm^{3+}$, relative to the VB maximum. The position of the occupied 4f band of $Eu^{3+}$ at –3.5 eV represents an irregularity. The irregularity originates probably in the fact that 3+ is not the typical oxidation state of the Eu atom and the formation of $Eu^{2+}$ is preferred. In $Ce^{3+}$ the band of unoccupied 4f states resides in the CB at ~4.7 eV. With increasing atomic number this band moves from the bottom of the CB into the gap. In $Eu^{3+}$ the band of empty spin-up 4f states resides at 0.7 eV. This continuous shift of the 4f band causes the decrease of the gap between the VB and the 4f band (cf. below).

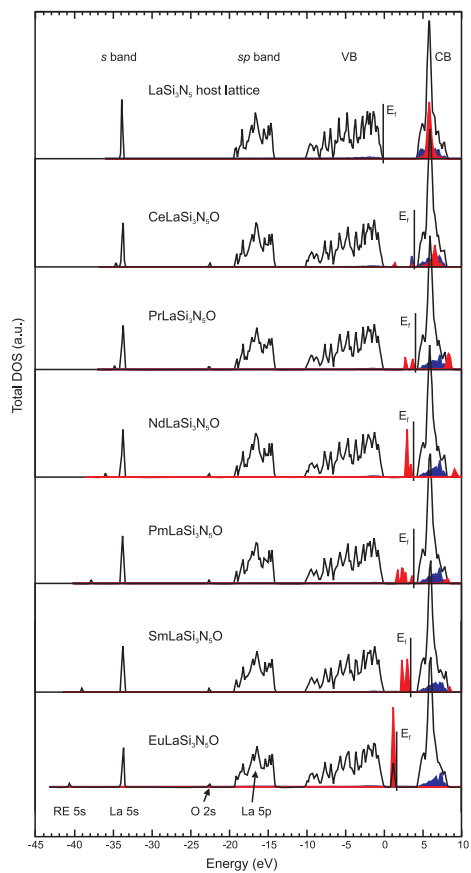

**Fig. 4.** Electronic structure of $LaSi_3N_5$ and $RE^{2+}$-doped $LaSi_3N_5$ (RE = Ce, Pr, Nd, Pm, Sm, Eu). Valence band maximum is set to zero, colors as in Fig. 3.

The electronic structure of the $LaSi_3N_5$ host structure doped with bivalent RE cations is displayed in Fig. 4. Compared with the DOS of trivalent compounds (Fig. 3) new features are visible only in the band gap originating in the RE 4f states. The change of the oxidation state causes a change of the electronic configuration of the RE cation from $4f^n$ to $4f^{n+1}$. The decreased atomic charge from 3 to 2 leads to a destabilization of 4f electrons. The DOS in Fig. 4 shows that a uniform shift of two 4f bands by ~6 eV occurs in the whole series of compounds. This shift causes that in compounds doped with bivalent RE cations the band of occupied 4f states is located directly in the gap. In the whole series of compounds the band of occupied states is completely filled. The upwards shift of the occupied 4f band causes also the shift of the Fermi level. While in compounds doped with trivalent RE cations the Fermi level is fixed at the edge of the valence band (Fig. 3), in compounds doped with bivalent cations the Fermi level resides at the edge of the band of 4f states (Fig. 4). With increasing atomic number the band of occupied 4f states moves towards more negative energies thus increasing the gap between the 4f band and the bottom of the CB (cf. below). In $Pr^{2+}$, $Nd^{2+}$, $Pm^{2+}$, and $Sm^{2+}$ compounds 4f spin-up states split into occupied and unoccupied states forming two bands separated by ~6 eV. In the $Pr^{2+}$-compound the occupied states accommodate three electrons (4f3) and unoccupied states correspond to four electrons. With increasing atomic number the number of 4f electrons increases and in the $Sm^{2+}$-compound the band of occupied states corresponds to the configuration $4f^6$ and the band of empty states corresponds to one electron. Different configurations are observed in $Ce^{2+}$- and $Eu^{2+}$-compounds. Two well separated small bands in the gap of the $Ce^{2+}$-compound correspond to the $4f^1 5d^1$ configuration. On the other side of the series in the $Eu^{2+}$-compound only a single band of completely occupied 4f spin-up states is formed ($4f^7$). Like in $RE^{3+}$ compounds 5d states in $RE^{2+}$ are spread over the whole width of the CB. The center of mass of the 5d states, however, is shifted to the upper part of the CB. Compared with corresponding states of trivalent RE cations 5d states in bivalent cations are destabilized. This is causes by the change of the oxidation state and the decrease of positive atomic charge. The destabilization of the 5d states in bivalent RE cations corresponds with the destabilization of 4f states (cf. above). The location of the 5d states in the CB, which is invariant of the lanthanide atom, agrees well with the position of 5d energy level of bivalent cations in the experimental energy level scheme by Dorenbos et al. [8]. With this location of the 5d states the $RE^{2+}$-doped compounds are not stable against the autoionization. This is consistent with the fact that d-f emission is never observed for divalent lanthanides on trivalent lattice sites [5].

*Energy level scheme*

The energy level scheme of trivalent and divalent lanthanides, derived from band positions in DOS diagrams (Figs. 3 and 4) is displayed in Fig. 5.

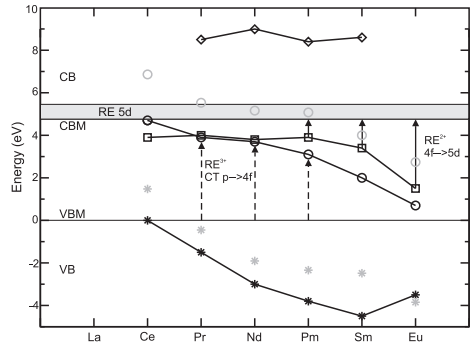

**Fig. 5.** Energy level scheme of divalent and trivalent lanthanides in $LaSi_3N_5$ host lattice. Asterisks (circles) show the position of occupied (unoccupied) 4f states of $RE^{3+}$, squares (diamonds) show occupied (unoccupied) 4f states of $RE^{2+}$. Gray symbols are experimental 4f ground states according to the energy level scheme in Ref. 8. Gray area indicates RE 5d states at the bottom of the CB. The dashed arrows indicate charge transfer energies to trivalent cations. Solid lines indicate f-d transitions within bivalent cations.

The scheme shows three separated series of 4f energy levels. The deepest levels are occupied $RE^{3+}$ 4f states located within the VB between 0 eV and -4.5 eV (cf. partial DOS in Fig. 3). The position deep in the VB and therefore larger stabilization of 4f states in trivalent cations, compared with $RE^{2+}$ and $RE^+$, is due to larger Coulomb attraction of 4f electrons in more positively charged cation. With increasing atomic number the location of occupied 4f states continuously decreases from 0 eV ($Ce^{3+}$) to -4.5 eV ($Sm^{3+}$). In $Eu^{3+}$ compound the position of occupied 4f states increases to –3.5 eV.

Second group of states are unoccupied 4f $RE^{3+}$ states, which are higher in energy than occupied 4f $RE^{3+}$ states. In the $Ce^{3+}$-compound empty 4f states reside at the bottom of the VB at 4.7 eV and with increasing atomic number the position of the band continuously decreases to 0.7 eV ($Eu^{3+}$). The location and the shape of the curve of empty 4f $RE^{3+}$ states is similar to the location and the shape of the curve of occupied $RE^{2+}$ states. Two curves cross each other and exhibit an overlap at Pr- and Nd-compounds.

Third series of 4f states are unoccupied states of $RE^{2+}$ cations with the configuration $5d^0 4f^n$, residing at ~8.4 eV (Fig. 5). No values are indicated for $Eu^{2+}$, $Ce^{2+}$, and $La^{2+}$. In the $Eu^{2+}$ cation the $4f^7$ configuration is formed with no empty spin-up states. The $Ce^{2+}$ and $La^{2+}$ cations form a configuration different from $5d^0 4f^n$. The electronic configuration of the former is $5d^1 4f^1$ and of the latter $5d^0 4f^0$.

The scheme compares reasonably well with the experimental scheme constructed for doped $LaSi_3N_5$ phosphors [8]. In Fig. 5 experimental 4f ground state energies are displayed as gray asterisks ($RE^{3+}$) and gray circles ($RE^{2+}$). In the experimental scheme the VB and the CB are separated by the gap of 5.2 eV, deduced from the excitation spectra of $Eu^{2+}$- and $Ce^{3+}$- doped $LaSi_3N_5$ [41, 42]. This band gap is a best estimate and includes the exciton binding energy, resulting in a band gap that is about 8% higher than the optical band gap of 4.8 eV [8]. The experimental scheme [8] shows occupied 4f and empty 5d energy levels of both trivalent and bivalent RE cations. Empty 4f energy levels are missing. In our theoretical scheme empty 5d levels are not drawn because 5d states are not collected in a narrow band but spread over the whole width of the CB. However, the center of mass of 5d states is in $RE^{3+}$ compounds deeper and in $RE^{2+}$ compounds higher in energy, in agreement with the separation of corresponding curves in experimental scheme [8]. It should be noted that in contrast to 4f states, 5d states are not collected in narrow bands because they are not completely nonbonding. The admixture of 5d states in the VB (Figs. 3 and 4) indicates that 5d states participate in bonding to some extent. The 5d states in the CB are therefore of slight antibonding character and therefore spread over a broad interval of energies. There is no reason to expect that 5d states could be collected in a narrow band of nonbonding states and localized below the CB. Not presented in the experimental scheme are empty 4f states, which are well localized in our calculated DOS (Figs. 3 and 4) below the CB and participate in CT p→f transitions (cf. dashed arrows in Fig. 5).

Compared with experimental data (gray symbols in Fig. 5) the curve of occupied 4f states of both trivalent and bivalent cations is shifted by ~1 eV towards negative energy. 4f states of $Ce^{3+}$ are therefore not located at ~1 eV above the VB maximum, like in the experimental scheme [8], but directly at the VB maximum. With increasing atomic number the calculated 4f energies decrease. The decrease reasonably compares with the decrease of experimental 4f energies. A different behavior is observed for $Eu^{3+}$. In the experimental scheme the $Eu^{3+}$ 4f band resides by ~1.2 eV deeper in the VB than that of $Sm^{3+}$. In the calculated scheme the $Eu^{3+}$ 4f band is higher in energy by ~1 eV than occupied $Sm^{3+}$ 4f band. This is due to much smaller separation of occupied and unoccupied 4f states in $Eu^{3+}$ than in other $RE^{3+}$ compounds (cf. Fig. 3).

In the experimental scheme $RE^{3+}$ 5d states are localized below the CB exhibiting a slight decrease from ~5 eV ($Ce^{3+}$) to ~4 eV ($Eu^{3+}$) [8]. In our scheme 5d states are inside the CB (gray area in Fig. 5). The experimental scheme shows a pronounced decrease of occupied $RE^{2+}$ 4f states from ~7 eV ($Ce^{2+}$) to ~2.7 eV ($Eu^{2+}$). The calculated energies of occupied $RE^{2+}$ 4f states exhibit a less pronounced decrease from ~3.9 eV ($Ce^{2+}$) to 1.5 eV ($Eu^{2+}$). The curve of occupied $RE^{2+}$ 4f states crosses the curve of empty $RE^{3+}$ 4f states, which shows much steeper decrease from 4.7 eV ($Ce^{3+}$) to 0.7 eV ($Eu^{3+}$). In experimental scheme empty $RE^{3+}$ 4f states are not identified. The experimental values of energies of unoccupied 5d $RE^{2+}$ states are almost independent of atomic number, located at ~5.5 eV at the bottom of the CB. This position agrees well with calculated location of 5d $RE^{2+}$ energies (gray area in Fig. 5). The calculated energies of unoccupied 4f $RE^{2+}$ states show only a slight variation. They are localized at the top of the CB at ~8.5 eV.

In the experimental scheme the separation of occupied 4f states of $RE^{3+}$ and $RE^{2+}$ varies between 6 and 7 eV [8] and towards the beginning of the lanthanide series the separation decreases to ~5 eV (Pr) and ~3.5 eV (Ce). Our calculated values show similar variation of the separation between 6 and 7 eV (Sm, Pm, Nd), decreased to 5.5 eV (Pr) and to 4.7 (Ce). Considerably decreased value of the separation of 4.2 eV, however, is observed for $Eu^{3+}$. This occurs because of too small splitting of occupied and unoccupied 4f states compared with other RE cations (Fig. 3).

In large band gap materials $RE^{3+}$ 4f energy levels of all lanthanides are located in the gap between VB and the CB. This is the case of $CaF_2$ exhibiting the gap larger than 12 eV [5]. In compounds with smaller gap the Fermi level shifts upwards in the energy level scheme. Upon the shift deep lying occupied $RE^{3+}$ 4f states remain localized below the Fermi level. E. g. in $YPO_4$ the band gap decreases to 9 eV, and the energy level of the occupied $Eu^{3+}$ 4f states is placed slightly below the Fermi level [5]. In LaSi3N5 the band gap decreases to 4.7 eV. The position of the Fermi level coincides with the energy of occupied 4f states of the $Ce^{3+}$ cation and 4f levels of all other $RE^{3+}$ cations are located below the Fermi level (Fig. 5). The fact that lines of energy levels of $RE^{3+}$ and $RE^{2+}$ cations located in different inorganic host lattices show similar shape means that the position of energy levels is driven by 4f and 5d states of the lanthanide cation. A systematic change in CT energy with type of lanthanide was revealed a long time ago for bromide complexes [5]. The CT band of $Sm^{3+}$ appears always 1.1±0.1 eV higher in energy than that of $Eu^{3+}$. Estimated from the experimental energy level scheme constructed for LaSi3N5 host lattice this value is ~1.26 eV [8]. In our scheme derived from ab initio calculations (Figs. 3 and 5) we observe the band of $Eu^{3+}$ and $Sm^{3+}$ 4f states at ~0.7 eV and ~2.0 eV above the Fermi level, respectively. In the LaSi3N5 host lattice the CT band of $Sm^{3+}$ is thus higher in energy than that of $Eu^{3+}$ by very similar value of ~1.30 eV.

*Band gaps*
Calculated band gaps of $LaSi_3N_5$: $RE^{3+}$ and $LaSi_3N_{5-x}O_x$: $RE^{2+}$ phosphors are shown in Fig. 6.

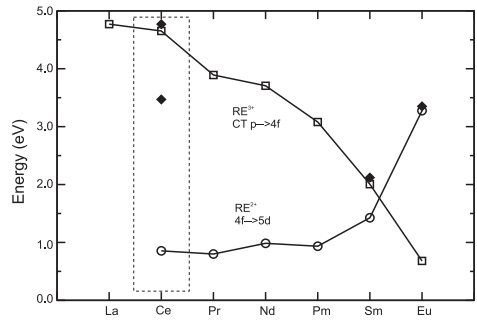

**Fig. 6.** Calculated band gaps of $RE^{3+}$-doped (squares) and $RE^{2+}$-doped (circles) $LaSi_3N_5$ compounds. Full diamonds indicate experimental data.

With the change of oxidation state of the doping RE atom a change of both the gap and the character of the transition occurs. In stoichiometric $LaSi_3N_5$ compound the maximum gap value of ~4.8 eV is calculated. This value compares well with the optical gap of 4.8 eV [43, 44]. The $LaSi_3N_5$:RE (RE = Pr, Nd, Pm, Sm, Eu) series shows a smooth change of the gap and the same character of the transition in the whole group of compounds. In compounds doped with the trivalent $RE^{3+}$ cation the minimum energy transition is the charge transfer p→4f and the gap decreases from ~3.9 eV (Pr) to ~0.7 eV (Eu). In the energy level scheme (Fig. 5) the CT p→4f transitions are indicated with dashed arrows. In compounds doped with bivalent cation $RE^{2+}$ the transition is 4f→5d and the gap increases from ~0.8 eV (Pr) to ~3.3 eV (Eu). In the energy level scheme (Fig. 5) the 4f→5d transitions are indicated with solid arrows. Experimental excitation energy of Sm-doped $LaSi_3N_5$ of 2.12 eV [29] fits the line of the $RE^{3+}$ doping (Fig. 6). On the contrary, the excitation energy of the Eu-doped $LaSi_3N_5$ of 3.35 eV fits the line of the $RE^{2+}$ doping [27, 41]. This is in agreement with the generally accepted opinion of the optical activity of the bivalent $Eu^{2+}$ cation. Ce-doped compounds exhibit deviations from the smooth behavior (Fig. 6). In the $Ce^{3+}$ compound the band of 4f states is not located inside the VB, but at the edge of the VB (cf. Fig. 3). The minimum energy transition is therefore not charge transfer p→4f, like in other compounds of the $RE^{3+}$-doped series, but 4f→5d, like in compounds of the $RE^{2+}$-doping. In the $Ce^{2+}$-doped compound the $4f^1\ 5d^1$ configuration is formed. The corresponding transition is therefore not 4f→5d, like in other $RE^{2+}$-doped compounds, but 5d→4f. Keeping in mind the tendency of Ce-compounds for increasing oxidation state to form tetravalent $Ce^{4+}$ cations it is unlikely that decreased oxidation state $Ce^{2+}$ can be formed. The second excitation band at ~4.8 eV of Ce-doped $LaSi_3N_5$ [28, 42, 43] fits the line of $RE^{3+}$-doping (Fig. 6). The first excitation band of Ce-doped compounds at ~3.5 eV [28, 42, 43], however, fits neither the curve of the $RE^{3+}$-doping, nor the curve of the $RE^{2+}$-doping. Is this an indication that some other doping, like formation of a non-integer mixed valency, occurs and makes prominent contribution to the optical activity of Ce-doped compounds?

**Conclusions**

We present the electronic structure of a series of RE-doped LaSi$_3$N$_5$ host lattice (RE= Ce, Pr, Nd, Pm, Sm, Eu). Following our recent work on synthesis, optical properties and model calculations of the LaSi3N5 host lattice doped with Eu [27], Ce [28] and Sm [29] in this work we perform calculations on Pr-, Nd-, and Pm-doped LaSi$_3$N$_5$ to complete the whole series of doped materials, in which the occupation of 4f energy level continuously increases from $4f^0$ to $4f^7$. In calculations the same model structure and same models of doping are used as in previous works [27–29] to simulate the doping with trivalent and bivalent RE cations. Our calculations use the hybrid screened HSE06 functional, which is expected to provide the correct position of nonbonding 4f-states.

The comparison of electronic structures of the series of compounds shows smooth changes in the position and the occupation of f-bands and provides the understanding of the doping mechanism. Upon the partial occupation the f-states the f band split into two bands. In the LaSi$_3$N$_5$ host lattice the band of occupied and unoccupied f-states are separated by ~6 eV. The position of the couple of f-bands strongly depends on the oxidation state of the RE cation. In LaSi$_3$N$_5$ the f-bands of the trivalent RE cation are shifted by ~6 eV towards negative energies than those of the bivalent RE cation. With increasing atomic number of the RE cation the position of the f-bands smoothly decreases.

In compounds doped with trivalent RE cation the band of occupied f-states is localized within the VB and the band of unoccupied f-states resides in the gap between the VB and the CB of the host lattice. The gap of the doped material is defined by the separation of the VB and the band of empty f-states (CT p→f transition). With increasing atomic number of the RE cation the f-band moves towards more negative energies and the gap continuously decreases from ~4.6 eV (Ce) to ~0.5 eV (Eu). In compounds doped with bivalent RE cation the band of occupied f-states resides in the gap between the VB and the CB and the band of unoccupied f-states is localized in the upper part of the CB. The gap is defined by the separation of the 4f band and the CB (f→d transition). With increasing atomic number the gap increases from ~0.8 eV (Ce) to ~3.3 eV (Eu). Empty RE 5d states are not as nonbonding as f-states. Because of the participation of 5d states in the bonding the unoccupied states are of slight antibonding character and spread over the whole width of the CB. In trivalent cation 5d states are more stabilized, localized more at the bottom of the CB, and in bivalent cation they are more destabilized, shifted to the upper part of the CB.

The energy level scheme constructed from the 4f and 5d band position as calculated using the hybrid functional reasonably agrees with the experimental scheme by Dorenbos et al. [8]. Some differences are observed at the beginning and the end of the series (Ce, Eu). For $Ce^{3+}$ the experimental scheme shows the gap of 3.5 eV, much smaller than the calculated value of ~4.7 eV. There are indications, however, that the gap of ~3.5 eV does not correspond to the trivalent cation, but to a non integer mixed valency between 3 and 4. The scheme shows the separation of 4f states in trivalent and bivalent Eu cation of ~6.3 eV. The separation calculated for the Eu-doped compounds is ~5.0 eV. The decreased separation is due to much smaller stabilization of occupied 4f states in $Eu^{3+}$ cation than in other $RE^{3+}$ cations. The stabilization of the occupied 4f states leads to a separation of occupied and unoccupied 4f states, which is typically ~6 eV. In $Eu^{3+}$-compound the separation is only 4.2 eV. Electronic reasons for much smaller stabilization of 4f states in $Eu^{3+}$-doped compound are unclear.

Our calculated scheme reproduces main features of the energy level scheme derived from spectral data. This means that hybrid functional HSE06 describes reasonably well 4f electrons in RE atoms and can be applied in the investigation of the electronic structure of RE compounds and host lattices doped with RE atoms.

**Acknowledgments**

This work was supported by the 7$^{th}$ FP Marie Curie Initial Training Network FUNEA (264875), Slovak projects VEGA 2/0112/14, ERDF project Center for applied research of new materials and technology transfer (ITMS code 26240220088), and the VASP project.

# TOC

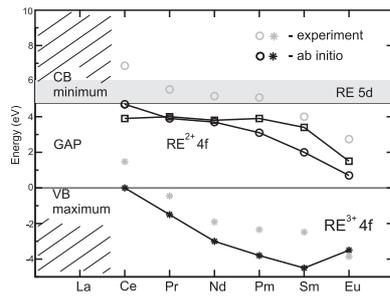

4f states of the lanthanide ion in the gap of the LaSi$_3$N$_5$ host lattice